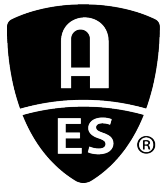

Audio Engineering Society

# Convention Paper 10186

# Exploring trends in audio mixes and masters: Insights from a dataset analysis

Angeliki Mourgela[1], Elio Quinton[1], Spyridon Bissas[1], Joshua D. Reiss[2], and David Ronan[1]

[1]*RoEx Ltd*
[2]*Queen Mary University of London*

Correspondence should be addressed to Angeliki Mourgela (`a.mourgela@roexaudio.com`)

**ABSTRACT**

We present an analysis of a dataset of audio metrics and aesthetic considerations about mixes and masters provided by the web platform MixCheck studio. The platform is designed for educational purposes, primarily targeting amateur music producers, and aimed at analysing their recordings prior to them being released. The analysis focuses on the following data points: integrated loudness, mono compatibility, presence of clipping and phase issues, compression and tonal profile across 30 user-specified genres. Both mixed (mixes) and mastered audio (masters) are included in the analysis, where mixes refer to the initial combination and balance of individual tracks, and masters refer to the final refined version optimized for distribution. Results show that loudness-related issues along with dynamics issues are the most prevalent, particularly in mastered audio. However mastered audio presents better results in compression than just mixed audio. Additionally, results show that mastered audio has a lower percentage of stereo field and phase issues.

## 1 Introduction

Recent advancements in audio production technology have significantly enhanced the capabilities of producers and audio engineers. However, practitioners continue to face numerous challenges that can negatively impact the quality of audio releases. Such challenges include managing loudness according to industry standards, ensuring phase coherence and mono compatibility, and achieving proper stereo balance and tonal profile. Furthermore, maintaining the right dynamic range without over-compression, combating listening fatigue, and navigating inadequate monitoring tools or poor room acoustics further complicate the process.

Various tools, including spectrum and phase analyzers, mono listening, and loudness metering, offer valuable assistance in assessing mixes and masters during the production phase [1, 2]. However despite the availability of established practices for monitoring and evaluating audio at this critical stage [3], persistent issues in released music underscore the necessity for a more thorough examination of trends in common audio quality concerns within the industry.

This paper presents the analysis of a comprehensive dataset comprising 218,109 entries with information on key parameters affecting audio quality, of which 67,838 were identified by users as mixes and 150,217 as masters. The dataset was derived from the online platform MixCheck Studio[1], a web-based audio analysis software, designed to provide actionable feedback and analytical insights on music tracks, highlighting the most prevalent issues in mixing and mastering, developed by RoEx.The platform was chosen due to the data availability which provided us with access to a diverse dataset of real world insights for mixes and masters.

---

[1]https://mixcheckstudio.roexaudio.com/

This research offers a detailed examination of the dataset, shedding light on the most frequent issues in audio production through analysis and visualization. By providing empirical data on common issues, this study aims to promote a better understanding of the current challenges faced in the industry.

The challenges explored in this study are:

- **Loudness levels:** Loudness monitoring is vital for ensuring that audio tracks are perceived as having an appropriate volume when played back on different devices and platforms[4]. Integrated loudness and true peak loudness measurements provide a standard for tracks to meet streaming service requirements.

- **Clipping:** Often characterized by perceptible distortion due to signal overload, and excessive loudness levels, directly impacts the dynamic range and perceived clarity of the audio [5]. Clipping occurs when the signal level exceeds the maximum limit of a system, leading to distortion. It can be minor, affecting only a few moments of the audio and sometimes even be imperceptible, or major, significantly impacting the overall sound quality.

- **Mono Compatibility:** Mono compatibility ensures that the essence and overall impact of a stereo mix are preserved when it is played back through a mono system.[6].

- **Phase Issues:** Can lead to a loss of audio fullness and definition, and present significant challenges in audio production. Phase issues can lead to comb filtering, which imparts a hollow and unnatural timbre to the sound. This can be particularly pronounced at lower frequencies[7].

- **Stereo Width:** Stereo field considerations involve the spatial distribution of sound, impacting how immersive and spacious a mix feels. A well-balanced stereo image enhances the listening experience by accurately representing the placement of instruments and elements within a mix, adding depth and movement[8].

- **Genre and perception based aesthetic considerations:** The study also examines the application of compression, more specifically looking for under/over or optimal compression in mixes and masters based on trends in the corresponding genres. Dynamic Range Compression (DRC) is used to balance the quietest and loudest parts of an audio signal, making it more consistent in volume [9, 10]. However, excessive compression can lead to a loss of dynamic expression, while similarly, excessive dynamics can make a song feel untamed and unbalanced. Additionally, the tonal profile is examined to show spectral characteristics and the amount of energy in specific bands of the tracks, based on perceptual metrics.

By analyzing these prevalent issues through the lens of a comprehensive dataset of real-life audio tracks, this paper not only underscores their commonality and impact on audio production but also sets the stage for discussing targeted interventions. The insights derived from this analysis aim to inform better practices in mixing and mastering, ultimately contributing to better education for amateur audio producers.

## 2 Methods

### 2.1 Dataset

The dataset consists of data obtained through the free online platform MixCheck studio. The platform allows the user to upload an audio track, select whether it is a mixed (i.e. not mastered) or a mastered track, as well as specify the track's genre. The platform then uses a series of audio signal processing analyses to determine common characteristics and issues that could be present in the audio, which are then presented to the user along with useful actionable feedback to help them improve their mix/master. The raw dataset can be accessed at: https://doi.org/10.5281/zenodo.13683186

The platform provides feedback on a range of standard and compatibility audio characteristics such as the track's sample rate, bit depth, integrated and true peak loudness, mono compatibility and phase issues, as well as more genre-based aesthetic characteristics aimed at helping the user improve the tonal, spatial and dynamic balance of their track. More specifically the aspects analysed by the platform are presented in Table 1 along with the type of analysis used for each aspect.

Table 1: Overview of the audio feature analysis methods employed by MixCheck Studio.

| Feature | Analysis Method |
|---|---|
| Bit depth, sample rate | Extracted from file |
| Integrated loudness | Calculated based on the ITU BS.1770 standard [11] |
| True peak | Calculated based on the ITU BS.1770 standard [11] |
| Compression | Calculated based on dynamic ranges in decibels (dB) of a signal by measuring 20 times the base-10 logarithm of the ratio between its maximum and mean amplitudes. Dynamic ranges are then compared to empirical average genre-specific levels found in the industry. |
| Stereo Width Balance | Calculated based on the inter-channel level difference. |
| Mono compatibility | Calculated by examining the correlation between the mid and side signals derived from the left and right stereo channels. |
| Tonal Profile | Calculated by dividing its frequency spectrum into defined bands (20-250, 250-2000, 2000-8000, 8000-20000) and comparing the energy within each band to thresholds relating to perceptual weighting. |
| Clipping | Calculated by counting how many samples exceed the digital full scale (0 dBFS) threshold, as well as evaluating the extent of clipping based on the number of clipped samples (over 10000 clipped samples signify major clipping) |
| Phase issues | Assesses the presence of phase issues between the left and right channels of a stereo audio signal by calculating the mean absolute phase difference and comparing it to a predefined threshold of 1.7 radians, set heuristically based on typical values observed in songs with noticeable out-of-phase issues. |

## 2.2 Data Analysis

For numerical descriptors we report aggregate statistics to summarize the central tendencies, dispersion, and distribution of numerical data, such as integrated loudness levels. For categorical data, such as musical genre, tonal profiles, and clipping, we used frequency distributions to understand the prevalence and distribution across different categories. Additionally, cross-tabulations were utilized to examine the relationship between variables, like clipping and musical genre. Another key part of our approach was the use of Cramér's V tests to examine the relationship between phase issues and the stereo field [12]. The statistical measure of Cramér's V, assesses the strength of association between two nominal variables, where a score of 0 denotes no association and 1 signifies perfect correlation. Finally, to explore the impact of dynamic range compression (DRC) during the mixing process on the tonal characteristics of music tracks, specifically focusing on bass and high frequencies, we employed Chi-square tests [13].

## 3 Results

### 3.1 Integrated and True Peak Loudness

Figure 1 displays a comparative analysis of integrated loudness and true peak loudness between audio mixes and masters. Plot (a) shows integrated loudness in LUFS (Loudness Units Full Scale), where the masters are generally louder than the mixes, as indicated by the density of the data leaning towards the 0 LUFS mark.This data clearly highlights the primary contribution of mastering, which is to increase the overall loudness of the recording. Mixes are more dispersed with a peak around -23 LUFS, while masters cluster tightly around -14 LUFS, which aligns with the target loudness levels required by most streaming platforms [14]. Plot (b) depicts true peak loudness in dBTP (decibels True Peak), with a high concentration of peaks at or just below 0 dBTP for masters, indicating a tendency towards maximizing loudness up to the threshold of clipping. Mixes, by contrast, show a wider distribution with true peak values commonly falling below -1 dBTP, signifying more conservative levels that avoid the risk of digital clipping. Both graphs highlight the pronounced difference in loudness handling between the mixing and mastering phases, with the increase in loudness achieved during the mastering process often resulting in true peak levels exceeding 0dB, thereby increasing the risk of clipping.

Approximately 79% of mastered tracks exceed -14 LUFS which is Spotify's loudness recommendation [14]. This suggests that a significant majority of mastered tracks are louder than what Spotify considers optimal, which means they may be turned down by Spo-tify's normalization process to match their target loudness level. Additionally, 91.55% of mastered tracks are louder than Apple Music's loudness recommendation of -16 LUFS [15]. This is an even larger majority compared to Spotify's standards, indicating that almost all mastered tracks are likely to be affected by Apple's normalization. Due to the fact that over half of the mixes are louder than -17.5 LUFS, it is hypothesized that they may not leave sufficient headroom for amplification during the mastering process, potentially resulting in a reduction in dynamic range. Finally, 10.24% of mixed tracks have a loudness lower than -23 LUFS, which suggests they may be too quiet for mastering and could potentially suffer from poor signal-to-noise ratios which directly affects limiting and compression.

### 3.2 Clipping

A significant portion of the mixes, at 68.58%, are free from clipping. In contrast, a lower percentage of masters, 42.53%, exhibit no clipping. Most genres show a higher incidence of major clipping at the mastering stage than at the mixing stage (see Fig 2), suggesting that the mastering process may introduce or amplify clipping. Minor clipping occurs when the number of samples that exceed the true peak level of 0dB is relatively low (<10000 clipped samples), indicating occasional instances of clipping that could be imperceptible. Major clipping, on the other hand, is identified when the number of samples exceeding the 0dB threshold is significantly higher (>10000 clipped samples), suggesting more frequent clipping throughout the audio track, that could potentially lead to audible distortion. The 'electronic' genre stands out with a notably high percentage of major clipping in both stages, while 'soul' shows a higher percentage of minor clipping during the mixing stage.

### 3.3 Mono Compatibility

Results from the mono compatibility analysis of mixes show a slightly higher percentage of compatibility issues with approximately 16.9% of the total mixes presenting with a lack of mono compatibility, compared to the masters, which is lower at around 12.0% of the total masters. This could suggest that mono compatibility is a more common concern in mixes than in masters. The reasons could be varied, but it could imply that the mastering process either resolves some of these mono compatibility issues or serves as a detection stage where these issues are identified and resolved by re-adjusting the mix.

### 3.4 Phase Issues

Results indicate a fairly close occurrence of phase issues between the two track categories, with mixes showing a slightly higher percentage at 16.3% compared to masters at 15.6%. This narrow gap suggests that phase issues are nearly as prevalent in masters as they are in mixes, hinting at the possibility that such issues are not significantly alleviated in the mastering process, or they may be introduced at similar rates at both stages of production.

### 3.5 Dynamic Range Compression (DRC)

Analysis revealed distinct trends in the application of dynamic range compression across mixes and masters. It was observed that mixes tend to have a higher incidence of under-compression, with 46.43% of mixes requiring more compression to reach optimal levels through empirical genre-based observations. In contrast, masters appeared to be more judiciously compressed, with a majority of 51.63% hitting the optimal mark. Conversely, when compression was less than ideal, mixes showed a slightly higher tendency at 17.00% compared to masters at 15.13%. Figure 3 shows the compression comparison between mixes and masters.

### 3.6 Stereo Field

There is a division between mixes with a wide (17.94%) and narrow (39.04%) stereo field, suggesting a variety in spatial imaging approaches. The stereo field for masters shows a preference for a wide (39.36%) over a narrow (16.45%) field, indicating a desire for spaciousness in the mastered tracks. There is also a small percentage of mono mixes and masters, which were excluded from the analysis.

### 3.7 Tonal profile

The graphs in Figure 5 and Figure 6 reflect the extent to which various musical genres deviate from a balanced tonal profile. Genres such as electronic, drum'n'bass, and techno demonstrate a pronounced emphasis on the bass frequencies. In contrast, genres like acoustic, folk, and blues lean more towards mid and high frequencies, suggesting a lack of balance in the lower frequencies.

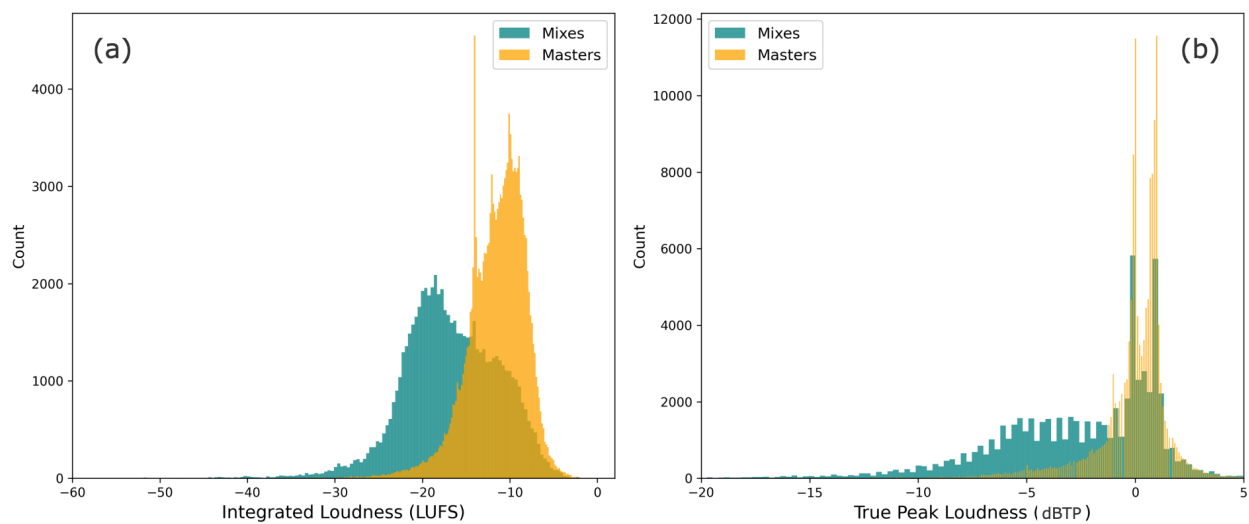

**Fig. 1:** Integrated (a) and true peak (b) loudness distribution for mixes and masters.

The notable instances where genres like orchestral and metal display peaks in the high-frequency range point to an excess of high-frequency elements.

### 3.8 Overall Issues

Table 2 presents a ranked comparison of the most common audio issues encountered in both mixes and masters. For mixes, the predominant issue is undercompression, followed by stereo field issues and excessive loudness. In contrast, masters are most frequently reported as being too loud, with clipping and overcompression as secondary concerns. Both mixes and masters share challenges with stereo field issues, though they appear at different ranks. Notably, issues like phase problems and lack of mono compatibility, while present in both, are less dominant, particularly in masters where the least common issue is being too quiet.

In addition to visualisations, relationships between all the aforementioned issues were investigated. More specifically, the relationship between stereo field type and incidence of phase issues, stereo field type and mono compatibility, as well as DRC and tonal profile were examined.

With regards to the relationship between stereo field type and incidence of phase issues, Cramer's V analysis for the mix data revealed a Cramér's V of 0.195, indicating a weak to moderate association between stereo field categorization and the presence of phase issues. This suggests a certain level of dependency, albeit not strong, pointing to some influence of stereo field setup on the likelihood of encountering phase issues. For masters, the Cramér's V slightly increased to approximately 0.213, again revealing a weak to moderate relationship between the stereo field and phase issues.

**Table 2:** Ranked issues for mixes and masters based on prevalence.

| Rank | Mixes | Masters |
|---|---|---|
| 1 | undercompression | too loud |
| 2 | stereo field issues | clipping |
| 3 | too loud | overcompression |
| 4 | clipping | stereo field issues |
| 5 | too quiet | undercompression |
| 6 | overcompression | phase issues |
| 7 | lack of mono compatibility | lack of mono compatibility |
| 8 | phase issues | too quiet |

With regards to the relationship between stereo field type and mono compatibility the results for mixes indicated a very weak association, with a Cramér's V of 0.059, while for masters, the association was similarly weak, evidenced by a Cramér's V of 0.073. These results underscore a very limited dependency between mono compatibility and stereo field categorization in the data.

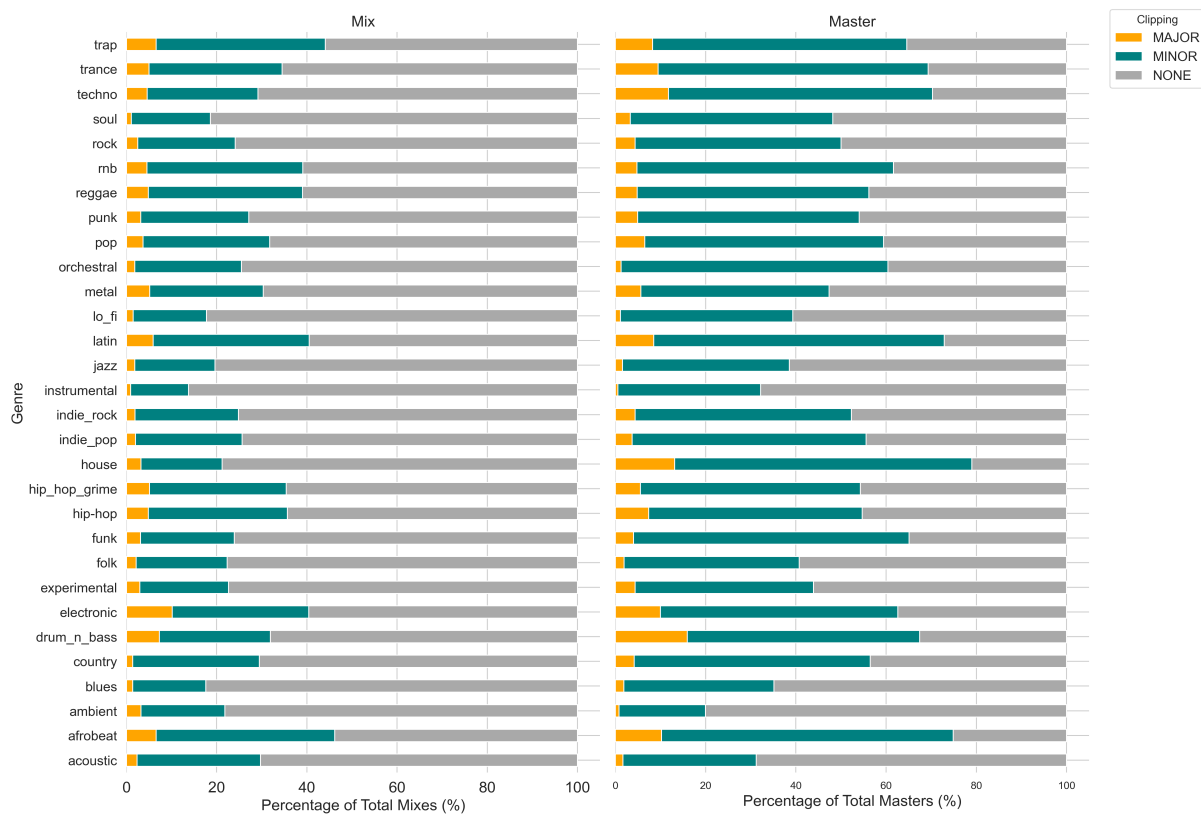

**Fig. 2:** Percentage of clipping types for mixes and masters, by music genre.

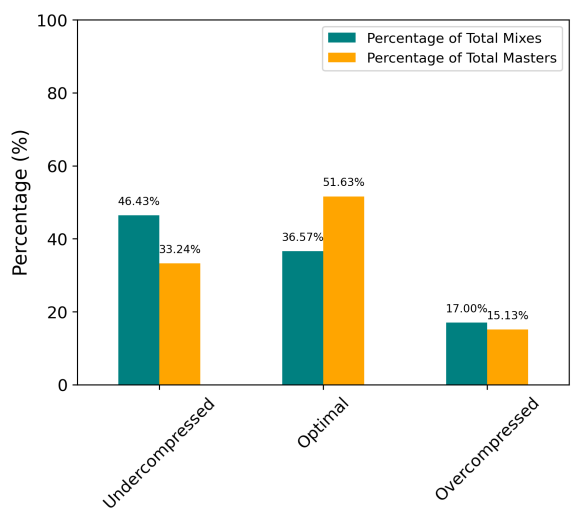

**Fig. 3:** Percentage of mixes and masters in the three compression categories.

Finally, as observed in Figures 5 and 6, low and high frequencies are most commonly characterised as having low energy in both mixes and masters. To examine the relationship between overcompression and low energy in these frequency bands a Chi-Square test was performed, to assess the statistical association between overcompression and frequency energy levels in the tonal profiles of mixes and masters.

For mixes, the analysis revealed statistically significant associations between the level of dynamic range compression applied during mixing and the energy levels in both frequency ranges. However, the direction and implications of these associations varied notably between bass and high frequencies, suggesting nuanced effects of mixing practices on the tonal balance of music tracks.

For bass frequencies, a tendency towards lower or medium bass frequencies was observed in tracks with overcompression, suggesting that overcompression during mixing could be linked with a reduction in the depth and power of bass sounds.

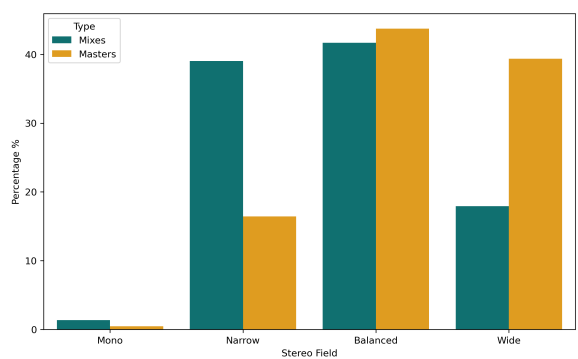

**Fig. 4:** Proportion of mixes and masters in the four stereo width categories.

For masters, our analysis revealed a highly significant association between the level of dynamic range compression (DRC) and the energy levels in both the bass and high frequencies of the tonal profile, with high values of chi-square statistics of approximately 1386.92 and 1263.23, and p-values nearing zero for both bass and high frequencies, respectively, indicating a robust statistical relationship. The direction of these associations, however, was different between the two frequency ranges. For bass frequencies, overcompressed tracks were more likely than expected to be associated with higher bass frequencies, suggesting that overcompression is associated with an emphasized bass response in the mastered tracks. Conversely, for high frequencies, overcompressed tracks were significantly more likely to exhibit lower high frequencies, potentially indicating a loss of clarity or brightness, characteristics often ascribed to the negative impacts of overcompression.

This contrasts with the earlier findings related to mixing, where overcompression was associated with lower bass frequencies. On the other hand, for high frequencies, overcompression during mastering was predominantly associated with lower high frequencies, reinforcing the hypothesis that overcompression can lead to a loss of clarity and brightness.

## 4  Discussion

Results from the dataset analysis highlight the prevalence of clipping, particularly in the masters. This indicates a tendency for producers to overuse amplification, which is evident in the correlation between excessive loudness and clipping. The genres that stand out for major clipping are electronic and trap in mixes and drum 'n' bass and house in masters. This could be attributed to genre-related trends, in addition to the influence of the "loudness war"[16].

Mono compatibility analysis demonstrated that masters present higher mono compatibility than mixes. This could be attributed to the fact that mastering is the last step in the audio production process prior to release, where usually the engineers will monitor the mixes in several different configurations and speakers. or employ mono listening to ensure coherence in playback. This allows them to identify compatibility issues and correct them or return the tracks to the mixing engineer to be re-mixed. Similar to mono compatibility mixes show a slightly higher percentage in the incidence of phase issues. This could also be attributed to interventions during the mastering stage.

Results from compression analysis demonstrated a tendency for undercompression in the mixes versus a slightly elevated incidence of overcompression in the masters. This could be attributed to the general practice of allowing for more dynamic range after mixing in order to allow for enough headroom to compress during mastering. These findings may suggest that compression is applied more effectively during the mastering stage, whereas mixes can be more prone to deviations from the optimal range, be it excess or insufficiency in compression. This can also be justified by the fact that mastering is the final stage of production before release therefore more careful precise interventions are applied to the dynamic range and usually by a different person so processing is approached with a fresh perspective.

Visualisations and analysis of the tonal profiles per genre show a tendency for better tonal control in the low-mid and high-mid frequencies compared to the low and high frequencies across mixes and masters. Overcompression has been shown to negatively impact high frequencies both in mixes and masters, however contrasting results were shown for the low frequencies. This could be attributed to either the way tonal profile and compression were measured in the platform since both rely on genre-based data, therefore poor genre reporting at the user end could affect the outcome of the analysis, or the use of multi-band compression at the mastering stage which would allow for a more targeted dynamic control in specific frequency bands.

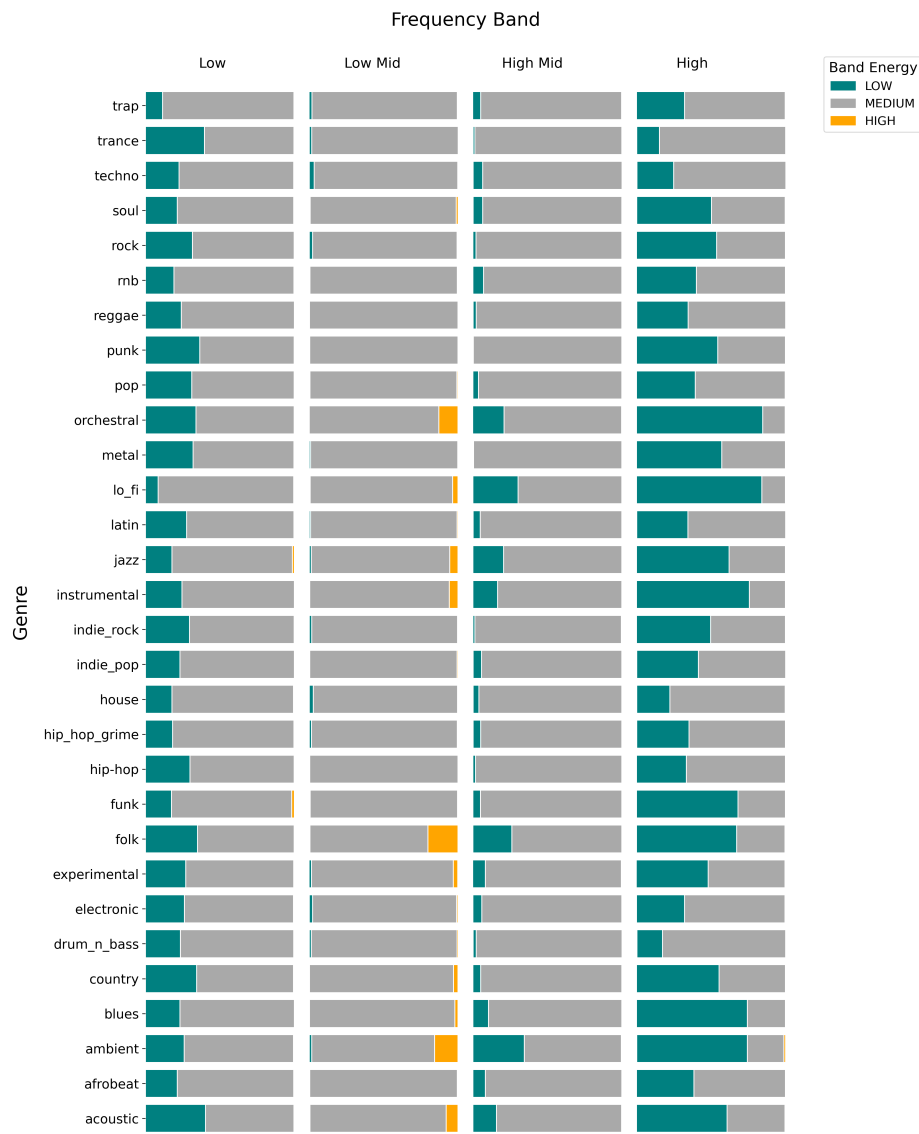

**Fig. 5:** Proportion of mixes categorized by low, medium, and high energy across the analyzed frequency bands.

Finally, an overall issue analysis suggests that dynamics processing, such as compression and loudness adjustments, along with clipping, are key areas that need careful attention during audio production to improve the quality of both mixes and masters.

This analysis offers insightful observations, yet it is important to note that it presents certain limitations. For instance, the analysis relies on user-reported information, such as the selection of musical genres and the distinction between mixes and masters. These aspects play an important role in shaping the dataset's various measurements. As a result, the reliability of outcomes is intertwined with the accuracy of this user-provided information.

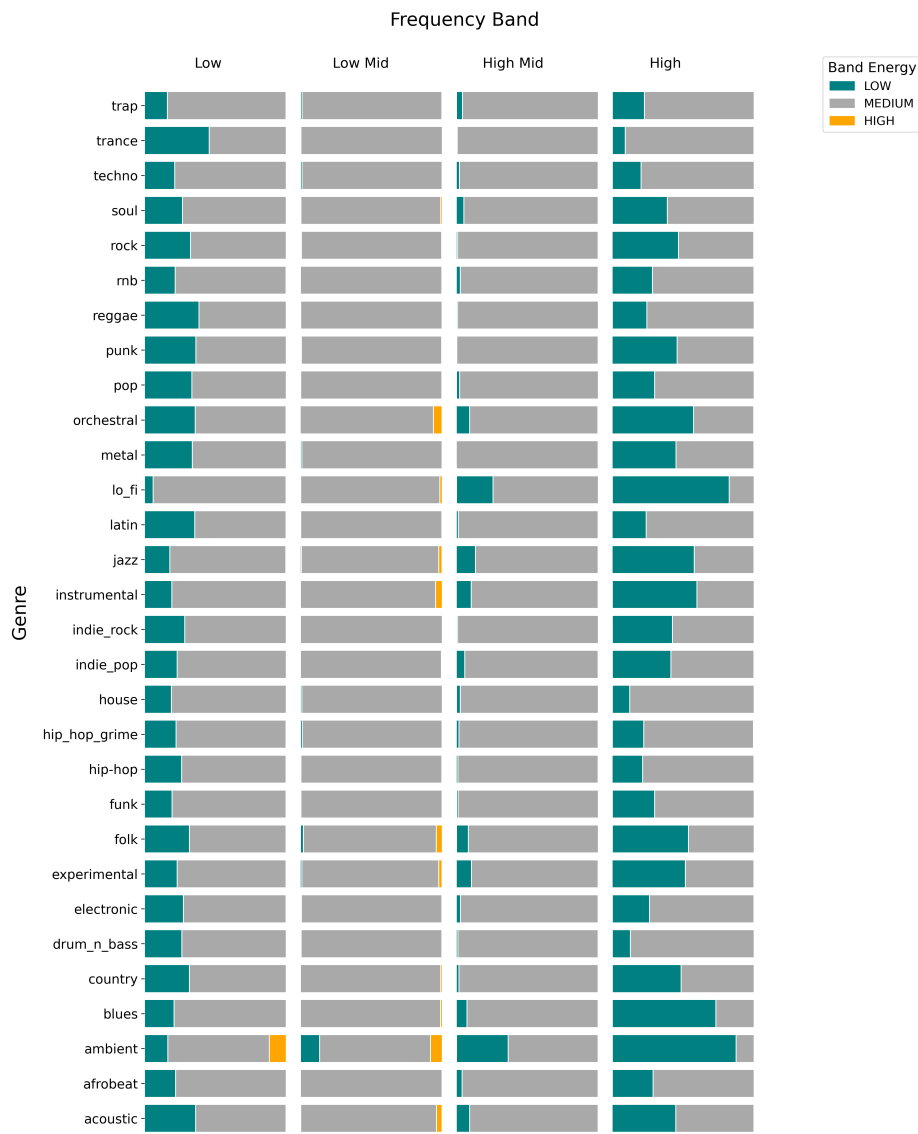

**Fig. 6:** Proportion of masters categorized by low, medium, and high energy across the analyzed frequency bands.

Moreover, the interpretations related to compression levels, tonal characteristics, and stereo field draw upon a combination of empirical understanding and current trends within the audio production and music industry. Such elements are not absolutes but rather reflect a consensus that evolves with artistic expression and technological advancements. As a result, these metrics should be regarded as informed estimates that encapsulate prevailing stylistic and aesthetic preferences rather than exact quantifications.

Our analysis revealed key trends and challenges in audio production, drawing on data from producers across a spectrum of experience levels, highlighting prevalent issues in both mixes and masters. Specifically, as detailed in Table 2, the most common problems in mixes include undercompression, stereo field limitations, and excessive loudness. Our study highlights several areas where users may need support to improve the quality of their mixes and masters. This support can come in various forms, including collaboration with industry experts as well as targeted training to help users develop the necessary skills to address specific audio issues effectively.

Finally, results from this analysis suggest an opportunity to develop knowledge engineering-based rules to enhance quality assurance (QA) in automatic mixing systems. Knowledge engineering in the context of audio production involves creating rules based on domain-specific knowledge to automate the decision-making processes of an automatic mixing system[17, 18, 19]. These rules can be derived from common patterns such as those identified in our dataset. Potential rules would include ensuring integrated loudness does not exceed specific thresholds, optimising dynamic range compression to maintain balance, detecting and effectively limiting samples that could exceed true peak levels, as well as analyzing and optimizing phase coherence while testing for mono compatibility. Additionally, evaluating the tonal profile for balanced spectral distribution can further refine the process and help such systems achieve better sound quality.

## 5 Summary

This study examined data from MixCheck Studio to pinpoint common problems in audio mixes and masters within 30 different genres. Its objective was to identify frequent issues related to audio quality, specifically targeting aspects like integrated loudness, clipping, phase issues, compression, and the tonal profile. The primary discoveries pointed to widespread issues with loudness and dynamics, particularly in masters, along with notable differences in clipping between mixes and masters. Masters were generally found to have superior mono compatibility and were compressed more effectively than mixes.

The findings underscore the importance of meticulous management of dynamics, loudness normalization, and clipping prevention in audio production. This research adds to the body of knowledge on the challenges faced in music production, offering insights that can improve audio production quality. Future studies could explore how technological advancements could further impact these production challenges as well as the possible remedies.